# LATTICE CONSTANTS AND EXPANSIVITIES OF GAS HYDRATES FROM 10 K UP TO THE STABILITY LIMIT

**Thomas C. Hansen[1], Andrzej Falenty[2], M. Mangir Murshed[2,°] and Werner F. Kuhs[2,*]**
[1]Institut Laue-Langevin, BP 156, 38042 Grenoble Cedex 9, FRANCE
[2]GZG, Abt. Kristallographie, Universität Göttingen, Goldschmidtstrasse 1, 37077 Göttingen, GERMANY

**ABSTRACT**
In a combination of neutron and synchrotron diffraction the lattice constants and expansivities of hydrogenated and deuterated $CH_4$-, $CO_2$-, Xe- (structure type I) and $N_2$-hydrate (structure type II) from 10 K up to the stability limit under pressure were established. Some important results emerge from our analysis: (1) Despite the larger guest-size of $CO_2$ as compared to methane, $CO_2$-hydrate has the smaller lattice constants at low temperatures which we ascribe to the larger attractive guest-host interaction of the $CO_2$-water system. (2) The expansivity of $CO_2$-hydrate is larger than for $CH_4$-hydrate which leads to larger lattice constants for the former at temperatures above ~ 150 K; this is likely due to the higher motional degrees of freedom of the $CO_2$ guest molecules (3) The cage filling does not affect significantly the lattice constants in $CH_4$- and $CO_2$-hydrate in contrast to Xe-hydrate for which the effect is quantitatively established. (4) Similar to ice Ih, the deuterated compounds have slightly larger lattice constants for all investigated systems which can be ascribed to the somewhat weaker H-bonding; the isotopic difference is smallest for the Xe-system, in which the large Xe atoms lead to an increase of averaged H-bond distances. (5) Compared to ice Ih the high temperature expansivities are about 50% larger; in contrast to ice Ih, there is no negative thermal expansion at low temperature.

**INTRODUCTION**
The thermal expansivity of gas hydrates has been investigated repeatedly and was summarized some years back in [1]. Unfortunately, rarely these studies extended to the temperature range above ice melting of central interest to geosciences and chemical engineering. This is because most studies were conducted at ambient pressure preventing experiments from reaching these temperatures without gas hydrate decomposition. Reliable information above 250 K is rare and mostly derived by making reference to more stable compounds like THF-or Xe-hydrate. To which extent information from these compounds can be transferred to natural gas hydrates remained questionable as Xe-hydrate was shown not to follow the trend of natural hydrates[1]. In the following we present experimental work on gas hydrate lattice constants and expansivities from 10 K up to the stability limit of the compound under pressure, i.e., covering the complete *p-T* range of interest for geosciences and chemical engineering.

Previous work has suggested that gas hydrates of structure type I and II (sI and sII, respectively) differ in their expansivities [1] with structure type I exhibiting the markedly higher values, yet it remains unclear whether deuterated and hydrogenated gas hydrates show any significant difference;

in [1] both types of hydrates were simply merged together without an explicit justification despite the fact that expansivities for hydrogenated and deuterated ice are different with the deuterated form having the higher values [2]. To clarify the magnitude of isotopic differences of lattice constants in gas hydrates is another purpose of this study.

Expansivity data are usually based on the diffraction measurement of lattice parameters. There are sometimes large deviations between different data sets published in literature, and the reasons are not always clear. The situation is illustrated in Figure 1 and Figure 2 for $CH_4$- and $CO_2$-hydrates. As an example we quote the large discrepancies between the works of [3-6]. For methane hydrate none of these agrees with any other, while the discrepancies are much larger than the respective estimated standard deviations. As the natural variation of lattice constants for methane hydrate was found to be fairly small with respect to the inter-experimental disagreement [3], the reason for this behavior can be suspected to be systematic errors in the data collection procedures of which we will discuss some of the most important ones in the following. The lattice constants quoted in literature can be affected by uncertainties of the measured lattice constants themselves but also by uncertainties of the temperature measurement. Concerning the achievable accuracy, temperature measurements in cryostats are definitely more reliable than data collections with open cold nitrogen flow systems. On the other hand, He-flow cryostats (e.g., the so-called "orange" cryostats frequently used in neutron diffraction experiments) can routinely achieve accuracies of a fraction of a degree.

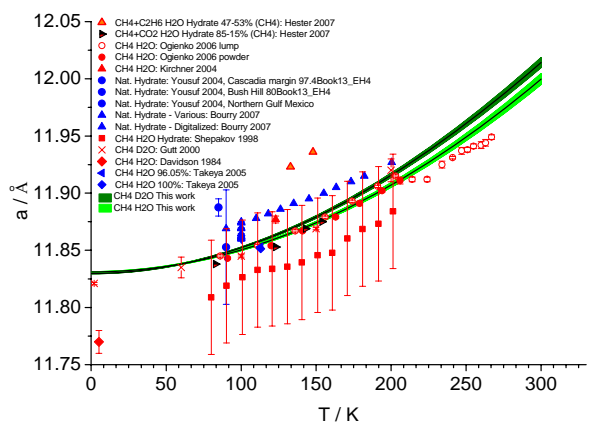

Figure 1. Temperature-dependent lattice constants of $CH_4$ and ($CH_4$,$CO_2$) mixed hydrates as taken from literature [1, 3, 5-11].

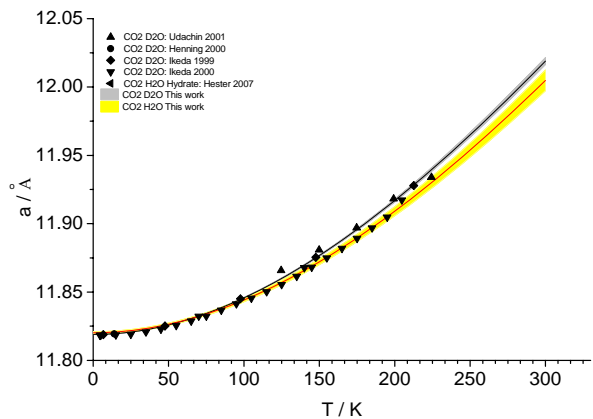

Figure 2. Lattice constants of $CO_2$ hydrates as taken from literature [1, 12-15].

Equally of concern is the vulnerability of profile refinement results of the lattice constants to the data collection techniques applied. Several types of errors can enter here relating to uncertainties of the wavelength, the zero point shift or the peak-shifts due to the umbrella effect (low-angle asymmetry); they all may affect the lattice constants refined in a full-pattern profile analysis via parameter correlations. Studies using the characteristic wavelength of a laboratory source can be considered to deliver quite reliable results, yet often this is combined with major uncertainties concerning the actual measurement temperatures (see above). Neutron time-of-flight studies are very reliable, concerning zero-point shift and wavelength, and parameter correlations are lower, in particular when considering data from the 90° detector bank. In all other cases careful calibrations need to be made to obtain reliable results, the evidence of which is sometimes lacking (in any case often not quoted) in published work on gas hydrates. In the following we try to overcome these difficulties by combining neutron and synchrotron X-ray diffraction data obtained on identical samples to take advantage of the unrivalled temperature-precision of orange cryostats used in neutron diffraction experiments and the unrivalled precision of lattice-constants obtained in high-resolution synchrotron powder diffraction experiments. Extensive cross-calibration was used to link both types of experiments.

**EXPERIMENTAL**
**Sample preparation and data collection**. The clathrate samples were prepared starting from $H_2O$ or $D_2O$ (99.9% deuterated) ice Ih, which was obtained by spraying water into liquid nitrogen re-sulting in ice spheres with a typical diameter of several tens of µm [16]. The production of deuterated ice took place in a glove box under dry nitrogen atmosphere to avoid isotopic exchange of the deuterated ice with atmospheric $H_2O$. The ice samples were placed into 7 mm diameter Al-vials or into larger PTFA-jars and inserted into pre-cooled pressure cells. The loaded cell was screwed onto the cryostat stick with its high-pressure capillary linked to the gas-handling system. Immediately after the cell was immersed into a cold bath and left there for about 10 minutes to achieve equilibration at the intended formation temperature (Table 1). Subsequently, the gas pressure was raised in a few seconds to the value of the chosen formation pressure. The pressure was monitored using Ashcroft pressure gauges, which were calibrated with a mechanical high-precision Heise manometer. An initial rapid pressure drop marked the onset of clathrate formation. To ensure a formation close to the target pressure the gas was re-adjusted manually over the whole reaction period of a few weeks (Table 1; the typical range of readjustments was 1-2 bars). At the end of the formation process the cell was cooled down in liquid $N_2$ until a rapid pressure drop in the reaction cell was registered. In the following 30-40 seconds the remaining gas was released and the cell opened. The samples were immediately quenched in liquid nitrogen and subsequently stored in a MVE Cryomoover Dewar flask.

The neutron diffraction experiments were performed at the high-flux powder diffractometer D20 at the High-flux reactor of the ILL in Grenoble set to a wavelength of $\lambda \approx 2.4182$ Å, using the (113) reflection of the vertically focusing germanium monochromator at a takeoff angle of 90° and a highly oriented pyrolytic graphite filter to eliminate higher harmonic wavelengths [17]. The samples were investigated over the temperature-range from 10 K up to ~ 270 - 290 K (the upper limit varying with pressure and guest molecule) by temperature ramping in steps of 30 K (5 K steps above 200 K), see Table 1. The pressure was set to values that guaranteed gas hydrate stability for the covered temperature range ($CH_4$-hydrate: 6 MPa; $CO_2$-hydrate: between 0.2 and 3.2 MPa; Xe-hydrate: between 0.5 and 1.0 MPa; $N_2$-hydrate: 15 MPa). Details on the high-pressure cell used can be found in [18]. The slight pressure readjustments for the Xe- and $CO_2$-hydrate samples were taken into account using the known compressibility data [19]; the results presented in the following are corrected for pressure-effects and refer to a pressure of 0.1 MPa. Deuterated as well as hydrogenated water frameworks were investigated; the hydrogenated sample presented a challenge due to the high incoherent background, in particular in the case of methane with its high hydrogen content of the guest molecules. To meet this challenge we had prepared samples with double-walled Al-vials (with a central rod of Al to prevent dense methane gas to be in the beam) as well as cans of 7 mm diameter (made of pure Al to avoid parasitic scattering from precipitates) completely filled with methane hydrate. Yet, it turned out that the data from the full cans had better signal-to-noise; thus in the further analysis only these data were considered.

| Hydrate | prepared at T/K | prepared at p/MPa | measured T-range/K |
|---|---|---|---|
| $CH_4$ $H_2O$ | 268.15 | 6.0 | 10 - 280 |
| $CH_4$ $D_2O$ | 271.15 | 6.0 | 10 - 280 |
| $CO_2$ $H_2O$ | 268.15 | 3.0 | 10 - 280 |
| $CO_2$ $D_2O$ | 271.15 | 3.0 | 10 - 280 |
| Xe $H_2O$ | 268.15 | 1.0 | 10 - 285 |
| Xe $D_2O$ | 271.15 | 1.0 | 10 - 160 |
| $N_2$ $H_2O$ | 258.15 | 15.0 | 10 - 270 |
| $N_2$ $D_2O$ | 258.15 | 15.0 | 10 - 270 |

Table 1. Details of sample preparation runs.

**Data analysis.** The analysis of the diffraction data was performed by sequential Rietveld refinement using FULLPROF [20]. The initial models for the refinement were taken from literature: for the hydrates of $CH_4$ and $CO_2$ [21], $N_2$ [22] and Xe [14]. The lattice constants, scale factors of the hydrate, a water ice impurity sometimes present ($D_2O$-$CH_4$, $H_2O$-$CO_2$, $H_2O/D_2O$-$N_2$), as well as a $3^{rd}$ to $6^{th}$ order background polynomial have been refined. In some cases, the resolution function ($D_2O$-$CO_2$), the occupancy of the guest molecule ($H_2O/D_2O$-$CO_2$, $D_2O$-$N_2$) or the coordinates and thermal motion of the framework atoms ($H_2O/D_2O$-$CO_2$, $H_2O/D_2O$-$N_2$, $H_2O$-Xe) could be refined too. While generally the refinement procedure went along smoothly, some problems were encountered, in particular for the hydrogen-rich samples, with correlations between the lattice constants and the zero-shift. The correlations lead to somewhat inconsistent results, which were overcome by fixing the zero-shift to the consistently refined value of deuterated samples. This was considered justified, as the set-up of the diffractometer remained unchanged upon a sample change. A further problem was encountered when trying to establish the absolute values of the refined lattice constants. Origi-nally, we were relying on an internal standard: the unreacted ice Ih remaining in the inner core of the gas hydrate grains. However, it turned out that this small amount of ice is likely to be under mechanical stress caused by the differences in expansivities of gas hydrates and ice. Therefore, we performed a separate cross-calibration with deuterated ice Ih [2]. Moreover, samples, identical to the ones used in the neutron diffraction experiments, were measured by synchrotron powder diffraction on ID31 at the ESRF in Grenoble to eliminate any possible systematic shift of lattice constants between the hydrogenated and deuterated forms which could arise in the neutron data e.g. as a consequence of correlations with asymmetry corrections.

**RESULTS**

The Rietveld refinements delivered a complete set of lattice constant data vs. temperature in the temperature range from 10 K up to the decomposition point together with their least-squares estimated standard deviations. These data form the basis of the subsequent analysis of the thermal expansion, which was performed within Mathematica [23]. In view of the complex nature of the guest molecules' motion with their many degrees of freedom, the various degrees of coupling of this motion to the host lattice [24, 25] we have used a polynomial approach for the description of the temperature dependency of expansivity to maintain sufficient flexibility of the mathematical model; terms up to the sixth order were needed for a satisfactory fit to the experimental data (any further order parameters delivered estimated standard deviations of the same order of magnitude as the additional parameters themselves). Considering the expansivity being zero at 0 K and being constant when approaching 0 K, the second and third term of the polynomial were tried to be fixed zero each. However, the fits were unsatisfactory in this way, and the latter of the two chosen constraints has been relaxed, as the measurement points were quite far away from 0 K anyway. In the Xe-case, even both constraints had to be relaxed. Expansivities were calculated by differentiation of the polynomial expressions for the lattice constants; the confidence intervals for the expansivity curves were obtained by a Monte-Carlo simulation procedure [26, 27] considering the variance-covariance matrix of the least-squares-estimated standard deviations of the polynomial coefficients. The results for the $T$-dependent lattice constants and the corresponding expansivities are shown in Figure 3 and 4, respectively.

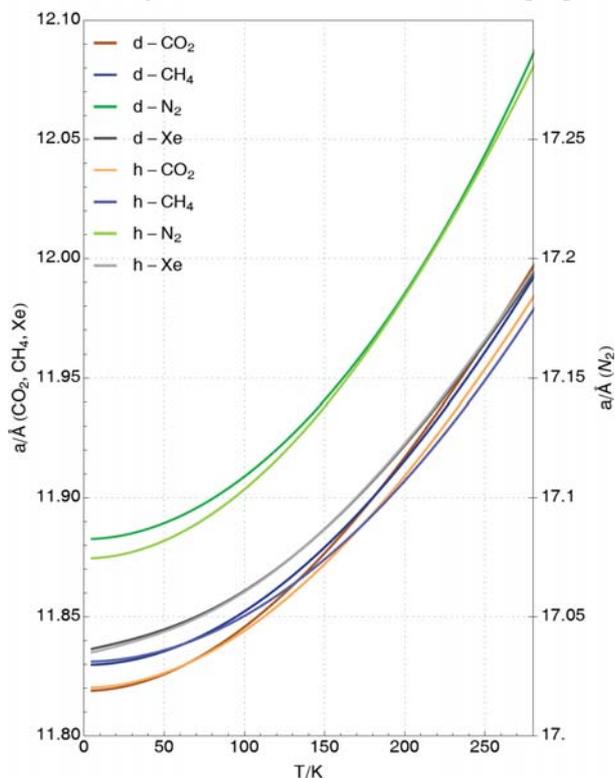

Figure 3. Lattice parameters as a function of temperature as described by least-squares-fitted polynomial expressions.

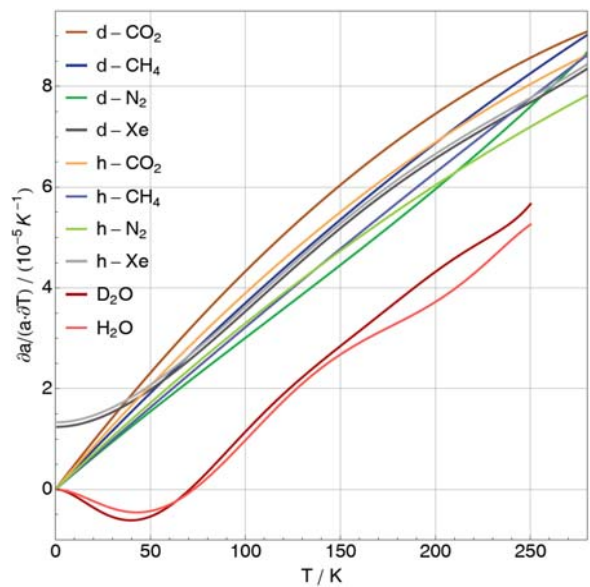

Figure 4. Linear expansivity vs. temperature for all investigated gas hydrates as well as for hydrogenated and deuterated ice Ih [2].

## DISCUSSION

Based on lattice constant data the thermal expansivities, $(\delta a/\delta T)/a$, of $CO_2$, $CH_4$, Xe and $N_2$-hydrates were obtained comprising also, for the first time, values for temperatures of geological interest. We cannot confirm the claimed large difference of expansivity for sI and sII gas hydrates, amounting to a difference by a factor of 1.5 at 200 K [1]; yet, indeed, $N_2$-clathrate has a somewhat smaller expansivity than the sI hydrates investigated. Also, the claimed smaller expansivity of Xe-hydrate as compared to other type I structures [1] is not really supported by our data. The results for $CO_2$-hydrate agree well with earlier neutron diffraction work [14], as shown in Figure 2. The very unsatisfactory situation for $CH_4$-hydrate (see Figure 1) is now clarified.

Reasons for the large inconsistencies in the methane case were discussed above and are likely to be related to calibration problems. The possible influence of cage-fillings on lattice constants was investigated as a possible further reason for the inconsistencies of published lattice constant data. The case was investigated for $CO_2$, $CH_4$ and Xe-hydrates. High-resolution synchrotron data were obtained for samples prepared at different p-T conditions with resulting differences in cage filling, in particular the filling of the small cages of their sI structures. These small cage fillings varied between 0.594 and 0.685 for $CO_2$-, 0.856 and 0.885 for $CH_4$-, and 0.808 and 0.909 for Xe-hydrate. The influence on the lattice constants for the $CO_2$- and $CH_4$-case was found to be very small and insignificant within our limit of accuracy. The situation is different for the Xe-case where for the higher cage filling the lattice constants are increased significantly by about 0.005 Å. This suggests that the large Xe atoms located in the small cages do indeed statically expand the water host lattice to some degree; indeed Xe-hydrate has the largest lattice constants of all sI hydrates investigated, in particular at lower temperatures.

The differences in lattice constants and expansivities between the hydrogenated and deuterated forms have the same sign as the differences observed in ice Ih, albeit with different magnitude. While $N_2$-hydrate is close to ice Ih in its isotope effect, the isotopic difference is reduced for $CO_2$- and $CH_4$-hydrate and quite small for Xe-hydrate. For ice Ih it was shown that the anomalous isotopic difference originates in coupling of quantum nuclear motion of H/D and H-bonding [28]; similar reasons can be expected to hold for gas hydrates. Thus, also for gas hydrates the deuterated form has somewhat weaker H-bonds translating into somewhat longer bonds. Earlier results have shown a similar result for $CH_4$-hydrate [19]. Moreover, also the expansivity of hydrogenated and deuterated form is different, with the deuterated form having about 10 - 15% higher values in the temperature range near ice melting for $CO_2$- and $CH_4$-hydrate. This may again be rationalized by the slightly weaker H-bonding between deuterated water molecules leading to a somewhat larger lattice anharmonicity. Thirdly, when one compares our results with the extrapolated T-dependency of expansivity in [1] two observations are in order: (1) the extrapolated numbers at 260 K are almost 20% too high for structure type I hydrates and at least 15% too low for structure type II hydrates. This means that the difference between hydrates of type I and II are considerably smaller than quoted in [1], at least for the cases investigated here. Rather, as shown in Fig. 4, some differences in detail are seen between different hydrates with different guests. Thus the extrapolation of the results obtained in [1] to higher temperatures is not supported by our data and a unified treatment for all clathrates as suggested in [1] is hardly justified. The complexity of interactions of the large-amplitude thermal rattling of the guest molecules with the host lattice [24, 25, 29, 30] prevents any simplified guest-independent approach. Detailed lattice-dynamical calculations will need to be performed to capture these interactions and may be compared with our results then. First attempts in this direction for the case of $CH_4$- and Xe-hydrate [31] and a comparison with our experimental results show that the differences between different guests are overestimated, as are – by a very considerable amount – the calculated expansivities.

It is established that lattice parameters influence in a subtle way the stability and cage fillings of gas hydrates [32]. It appears from our work that differences in lattice constants between $H_2O$ and $D_2O$ host lattices do exist in particular at higher temperatures. Thus, some differences between hydrogenated and deuterated gas hydrates in cage filling per hydration number can be expected to exist at *p-T* conditions relevant to geosciences and chemical engineering so that care must be taken when comparing results from neutron diffraction (usually done on deuterated material) with results obtained on hydrogenated samples or with calculations based on statistical thermodynamic theory – referring mostly to hydrogenated material. Our detailed results allow check earlier assumptions made by extrapolation and provide critical tests of estimates obtained from molecular dynamics simulations or lattice dynamical calculations. First estimates of lattice expansivity and compressibility have entered into improved thermodynamic prediction programs [32]; these entries could now be updated with our more detailed results, which may lead eventually to better predictions of thermodynamic properties.

**ACKNOWLEDGEMENTS**

The authors thank the Institut Laue-Langevin (ILL) for beam time and support as well as Paul Henry (ILL) for his advice in setting up the experiments on hydrogenated samples. Financial support was granted by BMBF in the framework of its GEOTECHNOLOGIEN and SUGAR-II programs as well as by DFG-grant Ku 920/11. The technical help of Eberhard Hensel and Heiner Bartels (both Göttingen) is gratefully acknowledged.